\documentclass[sigplan,10pt,nonacm,
screen]{acmart} 

\usepackage{xurl}

\AtBeginDocument{%
 }

\copyrightyear{2023}
\acmYear{2023}\setcopyright{rightsretained}
\acmConference[CPP '23]{Proceedings of the 12th ACM SIGPLAN
International Conference on Certified Programs and Proofs}{January
16--17, 2023}{Boston, MA, USA}
\acmBooktitle{Proceedings of the 12th ACM SIGPLAN International
Conference on Certified Programs and Proofs (CPP '23), January 16--17,
2023, Boston, MA, USA}
\acmDOI{10.1145/3573105.3575682}
\acmISBN{979-8-4007-0026-2/23/01}
\usepackage[sf,scaled=0.80]{noto}

\PassOptionsToPackage{hyphens}{url}
\usepackage{cleveref}
\usepackage{listings}

\usepackage[colorinlistoftodos]{todonotes}

\definecolor{keywordcolor}{rgb}{0.7, 0.1, 0.1}   % red
\definecolor{commentcolor}{rgb}{0.4, 0.4, 0.4}   % grey
\definecolor{errorcolor}{rgb}{0.9, 0.0, 0.0}     % bright red
\definecolor{symbolcolor}{rgb}{0.4, 0.4, 0.4}    % grey
\definecolor{sortcolor}{rgb}{0.1, 0.5, 0.1}      % green
\definecolor{stringcolor}{rgb}{0.31, 0.71, 0.14}      % green

\usepackage{xspace}
\newcommand{\mathlib}{\lstinline{mathlib}\xspace}
\newcommand{\QQ}{\mathbf{Q}}
\newcommand{\ZZ}{\mathbf{Z}}
\newcommand{\CC}{\mathbf{C}}
\newcommand{\NN}{\mathbf{N}}
\newcommand{\OK}{\mathcal{O}_K}
\DeclareMathOperator{\Norm}{Norm}
\DeclareMathOperator{\Cl}{Cl}
\DeclareMathOperator{\rk}{rk}

\begin{document}

\title{Formalized Class Group Computations and Integral Points on Mordell Elliptic Curves}

\author{Anne Baanen}
\authornote{Authors listed in alphabetical order}
\email{t.baanen@vu.nl}
\orcid{0000-0001-8497-3683}
\author{Alex J. Best}
\orcid{0000-0002-5741-674X}
\email{alex.j.best@gmail.com}
\author{Nirvana Coppola}
\orcid{0000-0002-8313-5984}
\email{nirvanac93@gmail.com}
\author{Sander R. Dahmen}
\orcid{0000-0002-0014-0789}
\email{s.r.dahmen@vu.nl}
\affiliation{%
 \institution{Vrije Universiteit Amsterdam}
 \streetaddress{De Boelelaan 1111}
 \city{Amsterdam}
 \country{The Netherlands}
 \postcode{1081 HV}
}

%% \renewcommand{\shortauthors}{Baanen, Best, Coppola, Dahmen}

\begin{abstract}
Diophantine equations are a popular and active area of research in number theory.
In this paper we consider Mordell equations, which are of the form $y^2=x^3+d$, where $d$ is a (given) nonzero integer number and all solutions in integers $x$ and $y$ have to be determined.
%Depending on $d$, such equations may have no integer solutions, or a positive (and finite) number of solutions.
One non-elementary approach for this problem is the resolution via descent and class groups.
Along these lines we formalized in Lean 3 the resolution of Mordell equations for several instances of $d<0$.
In order to achieve this, we needed to formalize several other theories from number theory that are interesting on their own as well, such as ideal norms, quadratic fields and rings, and explicit computations of the class number. Moreover, we introduced new computational tactics in order to carry out efficiently computations in quadratic rings and beyond.
\end{abstract}



%%  \includegraphics[width=\textwidth]{sampleteaser}
%%  \caption{Seattle Mariners at Spring Training, 2010.}
%%  \Description{Enjoying the baseball game from the third-base
%%  seats. Ichiro Suzuki preparing to bat.}

\maketitle

\section{Introduction}\label{sec:Intro}%\ab{try some different fonts}

The study of Diophantine equations forms a large part of modern algebraic number theory and arithmetic geometry \cite{cohen2007number}.
Compared to related areas of mathematics, relatively little of this topic has been covered in an interactive theorem prover.
This could be in part due to the lack of formalized prerequisites
necessary for work beyond elementary number theory,
such as the theory of algebraic number fields including class and unit groups.
Additionally, many Diophantine equations require explicit calculations for their effective resolution.
Such calculations are nowadays performed by computer algebra systems that have not been formally verified, like Mathematica, Magma, Pari/GP, and SageMath.

In this paper, we study the family of Diophantine equations known as Mordell equations,
\begin{equation}\label{eqn:Mordell}
 y^2=x^3+d, \quad x,y \in \ZZ\text,
\end{equation}
and formalize a full description of the solutions for various instances of the nonzero parameter $d \in \ZZ$ (Section~\ref{sec:Mordell}).
We classify the solutions, building on some fundamental algebraic number theory (Section~\ref{sec:Background}), using explicit calculations for quadratic number fields (Section~\ref{sec:QuadraticRings}) and class groups (Section~\ref{sec:class_group}).
Such calculations have not been carried out previously in a proof assistant.
The complexity of these arguments necessitates careful setup of the theory and developing abstractions to prove these results without extreme overhead.

We report on how the formalization is carried out and highlight three important outputs of our work:
First of all, there is the formalization of the explicit solution of some nontrivial Diophantine equations.
Secondly, we develop several other theories of independent interest, such as a theory of quadratic fields% (Section~\ref{sec:QuadraticRings})
, ideal norms, and class group computations.
Thirdly, we also set up computational structures and add tactics for performing calculations in extensions of commutative rings (Section~\ref{sec:computational-tactics}) and interacting with computer algebra systems (Section~\ref{sec:Sageify}).
Wherever possible, we aimed to work at a level of generality that applies beyond that needed for applications to the Mordell equations we considered.
Noteworthy difficulties that our formalization resolved are
developing general theory and calculating with specific numbers in the same setting of quadratic rings,
side conditions being resolved through computations in the typeclass system (Sections~\ref{sec:quad_ring_algebra} and \ref{sec:ring_of_integers}),
and avoiding more advanced geometry of numbers through an explicit method of calculating the class group (Section~\ref{sec:upper bound}).
We hope that this will pave the way for further formalization developments in (computational) algebraic number theory, and the explicit resolution of many other interesting Diophantine equations in particular.

The basic theory of Dedekind domains up to the finiteness of the class group of global fields was previously formalized \cite{baanen_formalization_2021} as part of the \mathlib library~\cite{mathlib} (using version 3 of the Lean theorem prover~\cite{lean-prover}).
We build on that work here, explicitly computing class groups of several quadratic number fields as part of our effort, and fleshing out the surrounding theory.
Full source code of our formalization is available online.\footnote{\url{https://github.com/lean-forward/class-group-and-mordell-equation}}

\section{Mathematical background}\label{sec:Background}

One of the basic tools we can use to investigate the solutions to a Diophantine equation, especially to show it has no solution, is the method of descent.
The main idea is to assume that a solution exists, and choose it to be minimal in some sense.
Then, transform this solution in order to `descend' to another solution to the equation that is smaller than the previous one, contradicting its minimality.
If the transformation can be unconditionally applied, the equation has no solutions;
otherwise, if the equation has solutions, it will lead to conditions on the solutions.

This method was used by Fermat to prove one instance of his famous Last Theorem, namely that the equation
\begin{equation}\label{eqn:Fermat4}
 x^4+y^4=z^4
 \end{equation}
 has no nonzero integer solutions. More precisely, Fermat proved (\cite[Section 1.6]{edwards-fermat}), in his notes on Diophantus' \emph{Arithmetica}, that there exists no right triangle with square area.
 This can be restated as the non-existence of a nonzero integer solution to the equation $ a^4-b^4=c^2 $, from which in turn the non-existence of a nonzero solution to \eqref{eqn:Fermat4} follows.
% This proof has been formalized in Lean\footnote{\url{https://github.com/leanprover-community/mathlib/blob/4e1eeebe63ac6d44585297e89c6e7ee5cbda487a/src/number\_theory/fermat4.lean}}.

The key step in the descent method is the application of the following property: if $a,b,c$ are three integer numbers and $n$ is a natural number such that $\gcd(a,b)=1$ and $ab=c^n$, then there exist $c_1,c_2 \in \ZZ$ such that $a=\pm c_1^n$ and $b=\pm c_2^n$. However, it is often convenient to use descent not over the ring $\ZZ$, but over some larger ring, e.g. the ring of integers of a quadratic number field.

We now recall some useful definitions. We assume some familiarity with basic ring and field theory, as described in many (undergraduate) texts, including~\cite{lang2005undergraduate, DummitFoote2004}.
A \emph{number field} $K$ is a finite extension of the field $\QQ$. It is a finite dimensional vector space over $\QQ$, and its dimension is called the \emph{degree} of $K$ (over $\QQ$).
Examples of number fields are $\QQ$ itself (of degree 1), $\QQ(\sqrt{5})=\{a+b\sqrt{5} : a,b \in \QQ\}$ and $\QQ(\sqrt{-5})=\{a+b \sqrt{-5} : a,b \in \QQ\}$ (both of degree $2$), 
and $\QQ(\sqrt[4]{17})=\{a+b\sqrt[4]{17}+c \sqrt[4]{17}^2 + d \sqrt[4]{17}^3 : a,b,c,d \in \QQ\}$ (of degree $4$).
%$\QQ(\alpha)=\{a+b\alpha+c \alpha^2 + d \alpha^3 : a,b,c,d \in \QQ\}$ (of degree $4$) where $\alpha$ is any complex root of the polynomial $x^4+x^3+2x^2+8x+64$ (the four different complex roots yield 2 distinct but isomorphic subfields of the complex numbers). 
%$\QQ(\sqrt[3]{2})=\{a+b\sqrt[3]{2}+c\sqrt[3]{2}^2 : a,b,c \in\QQ\}$ (of degree $3$), and $\QQ(\alpha)=\{a+b\alpha+c\alpha^2+d\alpha^3+e\alpha^4 : a,b,c,d,e \in \QQ\}$  (of degree $5$) where $\alpha$ is the real root of $x^5+2x+2$ (the other complex roots will yield isomorphic fields).
We will mostly use \emph{quadratic fields}, i.e. number fields of degree $2$ (Section \ref{sec:QuadraticRings}), as in the second and third example above. In general, any quadratic number field is isomorphic to a field of the form $\QQ(\sqrt{d})=\{a+b\sqrt{d} : a, b \in \QQ\}$ for some integer $d\not=1$ which is squarefree (i.e. not divisible by $p^2$ for a prime $p$).
Said differently, these quadratic number fields are constructed by adjoining a (complex) root $\alpha=\sqrt{d}$ of the polynomial $X^2-d$, which is irreducible over $\QQ$ by the conditions imposed on $d$.
This generalizes to arbitrary degree $n \in \ZZ_{>0}$ as follows.
For any polynomial of degree $n$ that is irreducible over $\QQ$, adjoining any one of its (complex) roots $\alpha$ to $\QQ$ will yield the degree $n$ number field $\QQ(\alpha)=\{a_0+a_1 \alpha + \ldots + a_{n-1} \alpha^{n-1} : a_0, a_1, \ldots, a_{n-1} \in \QQ\}$ (the $n$ different choices of $\alpha$ will all yield isomorphic fields). Conversely, any number field of degree $n$ will be isomorphic to such a field $\QQ(\alpha)$.

From an algebraic and arithmetic perspective, the (rational) integers $\ZZ$ constitute a particularly \lq nice\rq\ subring of its field of fractions, the rational numbers $\QQ$. Upon generalizing from $\QQ$ to an arbitrary number field $K$, the analogue of $\ZZ$ is the \emph{ring of integers} of a number field $K$, denoted $\OK$. It is defined as the integral closure of $\ZZ$ in $K$:
\begin{equation*}
\OK\coloneqq \{ x \in K : \exists f \in \ZZ[X] \text{ monic such that } f(x)=0 \},
\end{equation*}
where we recall that a (univariate) polynomial is called \emph{monic} if its leading coefficient is equal to $1$.
The fact that $\OK$ is indeed a ring follows for instance from general properties of integral closures (\cite[Section I.2]{Neukirch}).

Some examples of rings of integers are as follows.
The ring of integers of $\QQ$ is indeed $\ZZ$.
For $K= \QQ(\sqrt{d})$, values such as $d \in \{-5,-2,-1,2,3\}$ simply give $\OK=\ZZ[\sqrt{d}]=\{a+b \sqrt{d} : a,b \in \ZZ \}$,
but certain other values of $d$, e.g. $d \in \{-3,5\}$, give $\OK=\ZZ[\frac{1}{2}(1+\sqrt{d})]=\{a+b \frac{1}{2}(1+\sqrt{d}) : a,b \in \ZZ \}$.
That, for example, the ring of integers of $\QQ(\sqrt{-3})$ is strictly larger than $\ZZ[\sqrt{-3}]$ is apparent from the fact that $\frac{1}{2}(1+\sqrt{-3})$
(which is not an element of $\ZZ[\sqrt{-3}]$) is a root of the monic polynomial with integer coefficients $X^2-X+1$.
Finally, for $K=\QQ(\sqrt[4]{17})$ we have $\OK=\{a+b\sqrt[4]{17}+c\frac{1}{2}(\sqrt[4]{17}^2-1)+d \frac{1}{4}(\sqrt[4]{17}^3+\sqrt[4]{17}^2+\sqrt[4]{17}+1) : a,b,c,d \in \ZZ \}$, which cannot be written in the form $\ZZ[\alpha]$ for any $\alpha \in \OK$, illustrating the fact that determining rings of integers can become involved rather soon.

The \lq nice\rq\ properties alluded to earlier, can be made precise by the statement that the ring of integers $\OK$ of a number field $K$ is a so-called Dedekind domain \cite[Section 2]{baanen_formalization_2021}. While $\OK$ is not necessarily a unique factorization domain (UFD), nonzero ideals in $\OK$
(or more generally in any Dede\-kind domain) % ArXiv hyphenation hint
factor uniquely, up to the order of the factors, in (necessarily nonzero) prime ideals.

The key implication in the descent method still holds in general when $D$ is a UFD (replacing $\pm 1$ by a unit in $D$), for example when $D=\ZZ[i]$ is the Gaussian integers,
but is no longer applicable when $D$ is not a UFD, for example when $D=\ZZ[\sqrt{-5}]$.
However, when $D$ is a Dedekind domain such as the ring of integers of a number field,
we have a descent for ideals $I, J, L$ in $D$; namely the following implication holds for any natural number $n$:
\begin{equation}\label{eqn:Ideal_descent}
  \gcd(I,J)=1 \wedge I J=L^n \implies I=L_1^n \wedge J=L_2^n
\end{equation}
for certain ideals $L_1, L_2$ in $D$.
%\footnote{\url{https://github.com/leanprover-community/mathlib/blob/8c9342fd8949ec8421ed6649685d5233a19e0c08/src/algebra/gcd_monoid/basic.lean\#L530}}.

To return from ideals back to elements, the \emph{class group} of $D$, denoted $\Cl(D)$ is of crucial importance.
Recall that a fractional ideal $I$ of $D$ is a $D$-submodule of $K$ (the field of fractions of $D$) of the form $I=x J$ with $x \in K$ and $J$ an ideal of $D$.
Setting $x = 1$ shows every ideal $J \subseteq D$ is also a fractional ideal; for clarity we refer to this case as an \emph{integral ideal}.
% We call $I$ \emph{principal} if it is generated by a single element (in which case $I=xJ=x'D$ with $x'\in K$ the product of $x$ and a generator of $J$).
The class group is defined as the group of equivalence classes of nonzero fractional ideals of $D$ modulo nonzero fractional ideals of the form $\langle \alpha \rangle$, with ideal multiplication as the group operation.
Concretely, we say that fractional ideals $I \sim J$ if there exists a nonzero $\alpha \in K$ such that $ I = \langle \alpha \rangle J $~\cite[Section 2]{baanen_formalization_2021},
where we denote by $ \langle S \rangle$ the fractional ideal generated by the set $S$
(and by $\langle a_1, a_2, \dots, a_n\rangle$ the fractional ideal generated by the set $\{a_1, a_2, \dots, a_n\}$).
In particular, a nonzero (fractional or integral) ideal $I$ of $D$ represents the unit element in the class group of $D$ if and only if $I$ is principal.
We note that every equivalence class in the class group of $D$ has an integral ideal representing it,
and an equivalent definition of the class group could be given by taking appropriate equivalence classes ($I \sim J \Leftrightarrow aI=bJ$ for some nonzero $a,b \in D$) of the multiplicative monoid of nonzero integral ideals of $D$.

We consider the case at hand, where $D$ is the ring of integers of a number field,
for which the fundamental theorem holds that the class group of $D$ is finite,
as formalized by Baanen, Dahmen, Narayanan, and Nuccio~\cite{baanen_formalization_2021}.
The order of the class group of $D$, called the \emph{class number} of $D$ and denoted $h(D)\coloneqq\#\Cl(D)$, is a measure of the failure of unique factorization in $D$. In particular, $h(D)=1$ (i.e. the class group $\Cl(D)$ is trivial) if and only if $D$ is a UFD. So e.g. $h(\ZZ)=h(\ZZ[i])=1$.
It turns out that $\ZZ[\sqrt{-5}]$ is not  a UFD, since e.g. $I:=\langle 1+\sqrt{-5},2\rangle$ is a non-principal ideal, and in fact any non-principal (fractional) ideal of $\ZZ[\sqrt{-5}]$ is equivalent to $I$.
Hence a complete set of representatives for $\Cl(\ZZ[\sqrt{-5}])$ is given by $\{\langle 1 \rangle,I\}$ and $h(\ZZ[\sqrt{-5}])=2$.

An important property, needed for our approach to solving Mordell equations, which follows from the definition of the class group together with Lagrange's theorem in group theory, is that for ideals $I$ in $D$:
\begin{equation}\label{eqn:principality_from_power}
 \gcd(n,h(D))=1 \wedge I^n \text{ is principal} \implies I \text{ is principal}.
\end{equation}
Finally, we note that a number field $K$ determines uniquely its ring of integers $D$, and hence $\Cl(D)$ and $h(D)$. Therefore, the terminology \emph{class group of $K$}, denoted $\Cl(K)\coloneqq\Cl(D)$, and \emph{class number of $K$}, denoted $h(K)\coloneqq h(D)$, is also used.

The reader eagerly awaiting to see the above applied in the explicit context of the Mordell equation~\eqref{eqn:Mordell}, could already jump ahead to Theorem \ref{thm:Mordell_basic} and its proof sketch before proceeding with the next section.

\section{Quadratic rings and fields}\label{sec:QuadraticRings}

Crucial to the algebraic solution of Mordell equations is the ability to conveniently work with number fields and number rings such as $\QQ(\sqrt{-7})$ and $\ZZ\left[\frac{1}{2}(1 + \sqrt{5})\right]$,
    given by adjoining the root of a quadratic polynomial to a ring.
    There are several constructions that make the notion of a quadratic ring precise.
    One approach to adjoin a root of the polynomial $f \in R[X]$ to the domain $R$ is to take an algebraically closed (field) extension $K$ of (the fraction field of) $R$, which by definition contains a root $\alpha$ of $f$, and setting $R[\alpha]$ to be the smallest subring of $K$ containing each element of $R$ in addition to $\alpha$.
    In particular, this works well for number rings and number fields since in this setting we can always set $K$ to be the complex numbers.
    Alternatively, we can adjoin a root $\alpha$ to $R$ by setting $R[\alpha]$ to be the quotient ring $R[X] / \langle f\rangle$, where $\alpha$ is the equivalence class of $X$ mod $f$.
    Both constructions were previously available and used in \mathlib,
    finding use throughout the library in different settings.

    Additionally, \mathlib includes the structure \lstinline{zsqrtd} for the ring $\ZZ[\sqrt d]$, which is used to define the Gaussian integers, and prove some results related to Pell's equation.
    This suffers from the drawback that one can only adjoin square roots of integers to the integers, and hence cannot form quadratic fields, or the ring $\ZZ\left[\frac{1}{2}(1 + \sqrt{5})\right]$ using this structure.

    The general approaches described above suffer from the same two drawbacks for our purposes.
    The technical objection is that the definition of the new ring $R[\alpha]$ makes essential use of the ring structure on $R$, including the ring structure definition in the type $R[\alpha]$.
    Since Lean's unification algorithm performs poorly on complicated types, we want to avoid large definitions in types if possible.
    The more serious objection is that both of these two constructions are unsuited for efficient computations.
    Generally, \mathlib defines mathematical objects with little consideration of computability: computations in proofs are instead performed by constructing a proof term during tactic execution.
    Still, a tactic needs some useful data structure to perform its task, and a way to reflect expressions into this data type.
    Because there are no nontrivial decision procedures on the complex numbers, we cannot easily perform calculations on elements if we define $R[\alpha]$ to be a subring of the complex numbers.
    If we define $R[\alpha]$ as the quotient ring $R[X] / \langle f \rangle$, computation is possible by picking representatives of each equivalence class.
    Still, computing with $R[X] / \langle f\rangle$ will be impractical because Lean does not know that it can reduce the degree of polynomials to get a simpler representative.
    Moreover, \mathlib does not generally provide effective algorithms in its implementation of polynomial rings,
    meaning we cannot natively perform computations with the current implementation of polynomials anyway
    and we would have to reimplement arithmetic on polynomials in a tactic.

    Instead, for our project, following an established pattern (similar to e.g. defining complex numbers or quaternion algebras in \mathlib) we introduce a new, more specialized way to adjoin a root of a quadratic polynomial to a ring, which we name \lstinline{quad_ring}.
\begin{lstlisting}
structure quad_ring (R : Type*) (a b : R) :=
(b1 : R) (b2 : R)
\end{lstlisting}
    The code above defines a new type constructor \lstinline{quad_ring},
    such that an element of \lstinline{quad_ring R a b} is a tuple of two elements \lstinline{b1 b2 : R}, where \lstinline{R} is a commutative ring (actually, commutative semiring would also do).
    An element \lstinline{⟨b1, b2⟩} represents the element $b_1 + b_2 \alpha \in R[\alpha]$ where $\alpha$ satisfies the algebraic relation $\alpha^2=a\alpha+b$.
    We define a coercion to \lstinline{quad_ring},
    allowing us to implicitly consider an element \lstinline{x : R} also as an element of \lstinline{quad_ring R a b}.

    The fact that this type actually represents the ring $R[\alpha]$ follows from the definition of the ring structure, with some prior instances omitted this looks like:
\begin{lstlisting}
instance : comm_ring (quad_ring R a b) :=
{ add := λ z w, ⟨z.b1 + w.b1, z.b2 + w.b2⟩,
  mul := ⟨λ z w, ⟨z.1 * w.1 + z.2 * w.2 * b,
     z.2 * w.1 + z.1 * w.2 + z.2 * w.2 * a⟩⟩,
  neg := has_neg.neg, sub := has_sub.sub,
  one := 1, zero := 0,
  nsmul := (•), npow := npow_rec,
  nat_cast := coe, int_cast := coe, zsmul := (•),
  .. sorry } -- proofs omitted
\end{lstlisting}

We note that throughout this paper, we only omit proofs for sake of exposition; everything described here is fully formalized (i.e. \lstinline{sorry}-free) in our work.

    Beyond the basic ring operators such as addition and multiplication, we also define scalar multiplication with $\NN$ and $\ZZ$, and coercions from $\NN$ and $\ZZ$ to any \lstinline{quad_ring}.
    That we have to define these maps may be surprising since in fact we can define a unique ring homomorphism $\ZZ \to R$ generically for all rings $R$, which moreover gives rise to a unique $\ZZ$-module structure on $R$.
    Although we can prove these constructions are unique, Lean does not automatically consider them \emph{definitionally equal}.
    The \emph{forgetful inheritance} pattern used by \mathlib~\cite{Forgetful-inheritance,typeclasses} states that inheritance of structures,
    for example from \lstinline{comm_ring (quad_ring R a b)} to \lstinline{module ℤ (quad_ring R a b)},
    is not allowed to define new data.
    Instead, operations such as scalar multiplication must be copied literally from the parent, in this case the \lstinline{zsmul} field in \lstinline{comm_ring}.
    By specifying the operations ourselves in the \lstinline{comm_ring} instance, we obtain control over these definitional equalities.

    \subsection{Inclusion of $\ZZ[\alpha]$ into $\QQ(\alpha)$} \label{sec:quad_ring_algebra}

Within the quadratic field $\QQ(\alpha) = \QQ[x] / \langle f\rangle$ we have a subring $\ZZ[\alpha] = \ZZ[x] / \langle f\rangle$.
The informal description of a subring $A$ of a ring $R$ is a subset of $R$ closed under ring operations.
Here we define $A = \ZZ[\alpha]$ and $R = \QQ(\alpha)$ as quotients of two distinct rings, so $A$ cannot be a subset of $R$ in a straightforward way;
in Lean these would correspond to two different types, rather than a type \lstinline{R} and a term \lstinline{A : set R}.
In a similar way the integers $\ZZ$ are informally a subring of the rational numbers $\QQ$ but formally not a subset.
\mathlib instead prefers to translate the subring relation into an injective map: parameters \lstinline{(f : A →+* R) (hf : injective f)},
where \lstinline{f} is a ring homomorphism and \lstinline{hf} carries a proof of injectivity~\cite{baanen_formalization_2021}.
Subsets $A$ closed under ring operations fit this pattern by using the coercion map \lstinline{(↑)} that forgets \lstinline{x ∈ A}.

Moreover, given rings $A, R$, there is often a canonical choice for this inclusion $A \to R$.
The hypothesis that $A$ has a canonical map into $R$ is represented by a parameter \lstinline{[algebra A R]},
where \lstinline{algebra (A R : Type*) [comm_semiring A]} % ArXiv linebreak hack
\lstinline{[semiring R] : Type*} is a multiparameter typeclass depending on two types \lstinline{A, R}.
This provides a canonical homomorphism \lstinline{algebra_map A R : A →+* R};\\ % line break hack
thus a subring can be represented by a pair of parameters \lstinline{[algebra A R]} \lstinline{(h : injective (algebra_map A R))}. % line break hack

Here, the square brackets denote an instance parameter, a form of implicit parameter~\cite{typeclasses}.
Lean's elaborator automatically synthesizes a value for these parameters by considering declarations marked \lstinline{instance} in turn,
unifying its type with the parameter type and using the first instance that matches.
Instances can themselves have instance parameters, which are synthesized in the same way.
In Lean 3 synthesis is implemented as a depth-first search.
% while Lean 4 implements a more sophisticated algorithm that does not loop if the same goal repeats.
This search allows us to perform nontrivial computation during synthesis in the style of Prolog, as we discuss below.

In the case of quadratic rings, an \lstinline{algebra} instance would naïvely have the type \lstinline{algebra (quad_ring ℤ a b) (quad}\-\lstinline{_ring ℚ a b)}. % line break hack
However, this contains type errors since \lstinline{a b : ℤ} in the expression \lstinline{quad_ring ℤ a b} and \lstinline{a b : ℚ} in the expression \lstinline{quad_ring ℚ a b}.
Our first attempt was therefore to define the instance by applying the canonical inclusion of type \lstinline{ℤ → ℚ} to \lstinline{a b},
or more generally for any $A$-algebra $R$:
\begin{lstlisting}
instance algebra₁ [algebra A R] (a b : A) :
  algebra (quad_ring A a b) (quad_ring R
    (algebra_map A R a)
    (algebra_map A R b))
\end{lstlisting}
The verbosity of the expression \lstinline{quad_ring R (algebra_map A R a) (algebra_map A R b)} is unsatisfying,
and this gets worse when the coefficients are numerical expressions:
the inclusion of $\ZZ[\sqrt{5}]$ in $\QQ[\sqrt{5}]$
is denoted by \lstinline{algebra_map} \lstinline{(quad_ring ℤ 0 5) (quad_ring ℚ (algebra_map ℤ ℚ 0)}\\ \lstinline{(algebra_map ℤ ℚ 5))}.
Worse still, any result for \lstinline{quad}\-\lstinline{_ring ℚ 0 d} will need to be transported across the equality \lstinline{algebra}\-%ArXiv linebreak hack
\lstinline{_map ℤ ℚ 0 = 0}
before it applies to \lstinline{quad_ring ℚ} \lstinline{(algebra_}\-\lstinline{map ℤ ℚ 0) (algebra_map ℤ ℚ 5)}.

    To develop an instance without these deficiencies,
    we added the coefficients \lstinline{a' b' : R} as separate parameters to the \lstinline{algebra} instance,
    along with a pair of hypotheses asserting \lstinline{a' b'} are the image of \lstinline{a b} respectively,
    ensuring the map remains well-defined:
\begin{lstlisting}
def algebra₂ [algebra A R] (a b : A) (a' b' : R)
  (ha : algebra_map A R a = a')
  (hb : algebra_map A R b = b') :
  algebra (quad_ring A a b) (quad_ring R a' b')
\end{lstlisting}
    For numerical values of \lstinline{a b}, Lean's simplification procedure invoked by the \lstinline{simp} tactic suffices to prove the equality hypothesis,
    meaning we could define instances such as:
\begin{lstlisting}
instance algebra_sqrt_5 :
  algebra (quad_ring ℤ 0 5) (quad_ring ℚ 0 5) :=
algebra₂ ℤ ℚ 0 5 0 5 (by simp) (by simp)
\end{lstlisting}
    However, we could not declare \lstinline{algebra₂} to be an \lstinline{instance} globally,
    as the typeclass system is not able to supply values for the \lstinline{ha hb} parameters automatically.

    In general, to be able to infer an instance,
    all parameters to the instance must either occur in the type of the goal
    or be instance parameters themselves.
    In order to lift the equality hypotheses \lstinline{ha hb} to an instance parameter, we wrapped it in the \lstinline{fact} class~\cite{typeclasses}.
    This is a small wrapper class used in \mathlib that allows us to register certain proofs as instance parameters:
\begin{lstlisting}
class fact (p : Prop) : Prop := (out [] : p)
\end{lstlisting}
Using this, our algebra instance becomes (recalling that instance parameters are passed in [square brackets])
\begin{lstlisting}
instance algebra₃ [algebra A R]
  (a b : A) (a' b' : R)
  [fact (algebra_map A R a = a')]
  [fact (algebra_map A R b = b')] :
  algebra (quad_ring A a b) (quad_ring R a' b')
\end{lstlisting}
    Lean's expressive typeclass system with a Prolog-like synthesis allows it to perform computations.
    This means we can define a set of instances that automatically synthesize \lstinline{fact} instances of equality proofs by evaluation,
    or from a different perspective, we give a big-step operational semantics for expressions of the form \lstinline{algebra_map A R a = a'}.
    Computation steps that immediately terminate correspond to instances without further instance parameters:
\begin{lstlisting}
instance fact_eq_refl (a : R) : fact (a = a)
instance fact_map_zero (f : A →+* R) :
  fact (f 0 = 0)
instance fact_map_one (f : A →+* R) :
  fact (f 1 = 1)
\end{lstlisting}
	Computational steps that do not halt take an instance parameter representing the computation that still needs to be done, such as the \lstinline{h} parameter below:
\begin{lstlisting}
instance fact_map_neg (f : A →+* R)
  (a : A) (a' : R) [h : fact (f a = a')] :
  fact (f (-a) = -a')
\end{lstlisting}
    Finally, to handle numerical arguments we added two instances that evaluate the binary representation of these numerals
    given by the \lstinline{bit0} and \lstinline{bit1} functions:
\begin{lstlisting}
instance fact_map_bit0 (f : A →+* R)
  (a : A) (a' : R) [fact (f a = a')] :
  fact (f (bit0 a) = bit0 a')
instance fact_map_bit1 (f : A →+* R)
  (a : A) (a' : R) [fact (f a = a')] :
  fact (f (bit1 a) = bit1 a')
\end{lstlisting}

    The combination of the \lstinline{algebra₃} instance and the set of \lstinline{fact}s suffices for automatic synthesis of the desired instance:
\begin{lstlisting}
instance algebra_sqrt_5' :
  algebra (quad_ring ℤ 0 5) (quad_ring ℚ 0 5) :=
by apply_instance
\end{lstlisting}
Although Lean 3's instance synthesis algorithm lacks many of the refinements available in Prolog implementations, or indeed in the Lean 4 synthesis algorithm~\cite{lean-tabled-typeclasses},
we found that it was powerful enough to perform this limited set of computations within a reasonable time:
due to caching, the impact is restricted to a one-time cost of approximately 100 milliseconds per declaration
that contains goals of the form \lstinline{algebra (quad_ring ℤ 0 5) (quad_ring ℚ 0 5)}.
Costs\\ remain low also for larger numerals since the synthesis complexity is linear in the size of the binary representation.
We needed no tweaking of instances or other optimization to use this design in our project.

    \subsection{Rings of integers and Dedekind domains} \label{sec:ring_of_integers}

    After setting up the general theory, we specialize to rings of the form $R[\sqrt{d}]$,
in particular $\ZZ[\sqrt{d}]$ and $\QQ(\sqrt{d})=\QQ[\sqrt{d}]$ (we will use the notation \lstinline{ℤ[√d]} and \lstinline{ℚ(√d)} for these specific rings in code samples also).
    %and to the ring $\ZZ[\alpha]$ where $\alpha = \frac{1}{2}(1 + \sqrt{d})$.
    For a field $K$, we show that $K[\sqrt{d}]$ is a field if $d$ is not a square (otherwise, the ring would contain zero divisors).
    We first provide the field structure as a definition, not an instance, since the typeclass system is not set up to infer that a given $d$ is not a square.
\begin{lstlisting}
def field_of_not_square {K : Type*} [field K]
  {d : K} (hd : ¬ is_square d) :
  field (quad_ring K 0 d)
\end{lstlisting}
    Since it would be inconvenient to supply the field structure manually at each point it is required,
    we use the \lstinline{fact} construction to make propositions available to the typeclass system,
    so we can also define a \lstinline{field} instance:
\begin{lstlisting}
instance {K : Type*} [field K]
  {d : K} [hd : fact (¬ is_square d)] :
  field (quad_ring K 0 d) :=
field_of_not_square (fact.out hd)
\end{lstlisting}

    To show moreover that $\QQ(\sqrt{d})$ is the field of fractions of $\ZZ[\sqrt{d}]$,
    we again have to be careful to distinguish a number \lstinline{d : ℤ} from its image \lstinline{d' : ℚ},
    which we achieve using two separate variables and by adding a hypothesis \lstinline{[hdd : fact (algebra_map ℤ ℚ d = d')]}.
    We express that $K$ is the field of fractions of $R$ by registering a typeclass instance \lstinline{is_fraction_ring R K},
    which implies that $K$ is the smallest field that contains $R$;
    the standard proof of this fact was straightforward to formalize.

    Finally, we assume that $d$ is squarefree
    and show that $\ZZ[\sqrt{d}]$ is a Dedekind domain if $d \equiv 2$ or $d \equiv 3 \pmod 4$,
    and that $\ZZ[\alpha] = \ZZ[\frac{1}{2}(1 + \sqrt{d})]$ is a Dedekind domain if $d \equiv 1 \pmod 4$,
    by showing that these are the rings of integers of $\QQ(\sqrt d)$ under the given conditions.
    As the conditions suggest, the proof of this fact involves working in the ring $\ZZ / 4\ZZ$.
    The conditions provide a good example of the subtleties involved in dealing with coercions:
    for example we can write a condition as \lstinline{d % 4 = 3}, staying in $\ZZ$ and allowing direct computations to validate the condition,
    or we can write \lstinline{(↑d : zmod 4) = 3}, casting \lstinline{d} to an element of $\ZZ / 4\ZZ$ which is directly applicable when working in \lstinline{zmod 4}.
We choose to state this condition as \lstinline{d % 4 = 3} since existing proof automation can more easily check this condition;
whenever \lstinline{d} is cast to $\ZZ / 4\ZZ$ in a proof,
we can use the \lstinline{norm_cast} tactic~\cite{norm_cast} to prove that \lstinline{3 : ℤ} coerces to \lstinline{3 : zmod 4}.

    For the case $d \equiv 1 \pmod 4$, we instead choose $m : \ZZ$ so that $d = 4m + 1$,
    and write $\ZZ[\alpha]$ as \lstinline{quad_ring 1 m}.
    Again we provide an \lstinline{is_fraction_ring} instance, this time under the hypothesis \lstinline{[hdm : fact (d' = 4 * ↑m + 1)]}.
Although this instance could be generalized by replacing the coercion \lstinline{ℤ → ℚ} with \lstinline{algebra_map R K},
    we do not expect that this generality is worth trading in the higher ease of working with coercions.

    The fact that the discussed rings are Dedekind domains is a corollary of being the rings of integers of $\QQ(\sqrt{d})$.
    In other words, we have to show an element $x \in \QQ(\sqrt{d})$ is an element of $\ZZ[\sqrt{d}]$ or $\ZZ[\alpha]$ if and only if $x$ is integral:
    it is a root of some monic polynomial with integer coefficients.
    Since $\sqrt{d}$, $\alpha$ and each integer satisfy this property, and integrality is preserved by addition and multiplication,
    we can easily show each element of $\ZZ[\sqrt{d}]$ or $\ZZ[\alpha]$ is integral.
    We prove the converse by explicitly computing that the minimal polynomial (with coefficients in $\QQ$) of $a + b \sqrt{d}$ is $X^2 - 2aX + a^2 - d b^2$ (if $b \ne 0$);
    if $a + b \sqrt{d}$ is integral these coefficients are integers, since $\ZZ$ is a Euclidean domain with fraction field $\QQ$.
    If $d$ is not a square modulo $4$ (i.e. $d \in \{2, 3\} \pmod 4)$, this implies $b$ is an integer, from which it follows that $a$ is also an integer, giving $a + b \sqrt{d} \in \ZZ[\sqrt{d}]$.
    If $d \equiv 1 \pmod 4$, we can instead show there are integers $x, y$ such that $a = x + \frac{1}{2}y$ and $b = \frac{1}{2}y$,
    so that $a + b \sqrt{d} = x + y\frac{1}{2}(1 + \sqrt{d}) \in \ZZ[\alpha]$.
    The result is a pair of theorems
\begin{lstlisting}
theorem is_integral_closure_1 {m : ℤ} {d' : ℚ}
  [hdm : fact (d' = 4 * ↑m + 1)] :
  ¬is_square d' → squarefree (4 * m + 1) →
  is_integral_closure (quad_ring ℤ 1 m) ℤ
    ℚ(√d')
theorem is_integral_closure_23 {d : ℤ} {d' : ℚ}
  [hdd' : fact (algebra_map ℤ ℚ d = d')] :
  d % 4 = 2 ∨ d % 4 = 3 → squarefree d →
  is_integral_closure ℤ[√d] ℤ ℚ(√d')
\end{lstlisting}
giving models for the rings of integers of $\QQ(\sqrt d)$.

\section{Class group calculations}\label{sec:class_group}

A classical result, formalized by Baanen, Dahmen, Narayanan, and Nuccio~\cite{baanen_formalization_2021}, is the finiteness of the class group of a number field.
We have built on this formalization to compute some nontrivial class groups.
Although the existing formalization could be made effective by supplying an explicit description of the ring of integers,
it provides very weak bounds that would result in unnecessary work,
so we chose to redo the computations explicitly.
In general, there are two basic ingredients to get to the class group of a number field $K$ with ring of integers $\mathcal{O}_K$. First of all, one needs to get to a finite generating set for the class group. A well-known
classical result that leads to this is the Minkowski bound, as briefly discussed at the end of this section.
However, we have followed a more direct approach via generalized division with remainder, as explained in the subsection below.
Finally, given a finite generating set, one needs to establish the multiplicative structure to fully determine the group. In our current approach we focus on small, including nontrivial, examples.

\subsection{Upper bound} \label{sec:upper bound}

We will build on the formalization in~\cite[Section 8.2]{baanen_formalization_2021} about the finiteness of the class group of a number field $K$.
Our starting point is the following generalization of a theorem formalized in op. cit.

\begin{theorem}\label{thm main class group divisibility}
Let $I$ be a nonzero ideal of $\OK$,  let $M$ be a finite set of nonzero integers, and assume that
for all $a,b \in \mathcal{O}_K$ with $b\not= 0$ there exist $q \in \OK$ and $r \in M$ with
\begin{align}
  |\Norm_{\OK/\ZZ}(ra-qb)| < |\Norm_{\OK/\ZZ}(b)|. \label{eqn:div-with-remainder}
\end{align}
Then there exists an ideal $J$ of $\OK$ such that
\[I \sim J \text{ and } J \,\left|\, \middle\langle \prod_{m \in M} m \right\rangle.\]
\end{theorem}

We note that instead of the ideal generated by the product over all elements in $M$, we could take the ideal generated by any nonzero common multiple of the elements of $M$ (e.g. their least common multiple).
But, when focusing on generators, we can use unique factorization of ideals to restrict $J$ to prime ideals,
so by simply taking the product over all elements in $M$ we do not introduce any more prime ideals.
We formalized the above statement in Lean,
generalizing $\ZZ$ to any Euclidean domain \lstinline{R} with an absolute value operation \lstinline{abv}
and $\OK$ to any Dedekind domain \lstinline{S} such that \lstinline{L = Frac(S)} is a finite extension of \lstinline{K = Frac(R)}.

\begin{lstlisting}
theorem exists_mk0_eq_mk0' (I : (ideal S)⁰)
  (M : finset R) :
  (∀ m ∈ M, algebra_map R S m ≠ 0) →
  (∀ (a : S) (b ≠ 0) → (∃ (q : S) (r ∈ M),
    abv (norm R (r • a - q * b)) <
      abv (norm R b))) →
  ∃ (J : (ideal S)⁰), mk0 L I = mk0 L J ∧
    algebra_map _ _ (∏ m in M, m) ∈ J
\end{lstlisting}
Here \lstinline{mk0 L} is the homomorphism sending a nonzero ideal \lstinline{I : (ideal S)⁰} to its equivalence class in the class group.

To establish the assumption for the generalized division with remainder given in Theorem \ref{thm main class group divisibility},
it is practical to translate the norm estimates~\eqref{eqn:div-with-remainder} in the ring of integers to norm estimates in the field of fractions, dividing both sides by a factor \lstinline{b}.
\begin{lstlisting}
lemma exists_lt_norm_iff_exists_le_one
  [is_integral_closure S R L]
  (M : finset R) :
  (∀ (a : S) (b ≠ 0), ∃ (q : S) (r ∈ M),
    abv (norm R (r • a - q * b)) <
      abv (norm R b)) ↔
  (∀ (γ : L), ∃ (q : S) (r ∈ M),
    abv (norm K (r • γ - q • 1)) < 1)
\end{lstlisting}
In the proof we spent the most work tediously showing \lstinline{b} is nonzero iff its image in \lstinline{L} is nonzero, iff its norm over \lstinline{R} is nonzero, iff its norm over \lstinline{K} is nonzero,
to justify the idea of dividing by \lstinline{b} on both sides.

We now move to concrete corollaries of the above results.
We consider quadratic number fields $K=\QQ(\sqrt{d})$ where $d \in \ZZ$ and assume that $d\equiv 2 \text{ or } 3 \pmod{4}$ and that $d$ is squarefree (unless explicitly indicated otherwise),
This is specified in Lean using the typeclass assumptions \lstinline{[fact (d % 4 = 2 ∨ d % 4 = 3)]} and \lstinline{[fact} \lstinline{(square}\-\lstinline{free d)]} respectively, as these are needed for the Dedekind domain instance anyway.
%In our examples $d$ will in fact always be negative.
To get to the class group results for $d \in \{-1,-2,-5,-6 \}$, we perform the necessary norm estimates for such $\gamma \in \QQ(\sqrt{d})$ with the set $M$ taken to be $\{1,2\}$:

\begin{lstlisting}
theorem sqrt_neg.exists_q_r
  [fact (algebra_map ℤ ℚ d = d')]
  (γ : ℚ(√d')) : -6 ≤ d → d ≤ 0 →
  ∃ (q : ℤ[√d]) (r ∈ {1, 2}),
    abs (norm ℚ (r • γ - q • 1)) < 1
\end{lstlisting}

The proof is elementary and consists essentially of case distinctions depending on the components of $\gamma$, which are rational numbers, and establishing inequalities for these components.
After we set up estimates for the nonlinear parts of the inequalities,
the tactic \lstinline{linarith} could discharge the goals.

Combining this theorem above with the two earlier mentioned results
now shows that for $d\in \{-1,-2,-5,-6\}$ any nonzero ideal $I$ in $\ZZ[\sqrt{d}]$ is equivalent in the class group to an ideal $J$ dividing $\langle 1 \rangle$ or $\langle 2 \rangle$.
The next task is to factor $\langle 2 \rangle$ into prime ideals.
To this end, we explicitly define a nonzero ideal \lstinline{sqrt_2 d} so that \lstinline!(sqrt_2 d)^2! is the ideal generated by $2$,
and we show that this ideal is irreducible.
The definition of \lstinline{sqrt_2 d} depends on the remainder of $d$ modulo $4$.
\begin{lstlisting}
def sqrt_2 [fact (d % 4 = 2 ∨ d % 4 = 3)] :
  (ideal ℤ[√d])⁰ :=
if d % 4 = 2 then
  ⟨ideal.span {(⟨0, 1⟩ : ℤ[√d]), 2},
   sorry⟩ -- Proof of nonzeroness omitted
else
  ⟨ideal.span {(⟨1, 1⟩ : ℤ[√d]), 2},
   sorry⟩ -- Proof of nonzeroness omitted
\end{lstlisting}
We check that \lstinline!span {2}! factors as \lstinline{(sqrt_2 d)^2} in both cases by explicitly showing the set of generators of each ideal is contained in the other ideal.
Checking this equality was far more demanding than it would appear based on the simplicity of the informal argument, and the formalization of each equality takes up more than 20 lines of code.
This is because there are many routine divisibilities and explicit numerical conditions to be checked that currently have to be guided by hand.
This is an area in which improved tactics or better abstractions should be developed to reduce the work in carrying out similar calculations in a proof assistant.
% in the future.

To show that \lstinline{sqrt_2 d} is a prime ideal, we consider its norm.
In algebraic number theory the norm of an integral ideal can be taken as an integer, which is the \emph{absolute ideal norm},
or as an ideal of a subfield, giving the \emph{relative ideal norm}.
We formalized the absolute ideal norm for Dedekind domains $R$ as a multiplicative homomorphism, which is defined for an ideal $I$ of $R$ as $N(I) = \left|R / I\right|$ if the quotient $R/I$ is finite, and $0$ otherwise.
We prove multiplicativity in two steps: first we show $N(IJ) = N(I) N(J)$ when $I$ and $J$ are coprime,
then we show $N(P^i) = N(P)^i$ when $P$ is a prime ideal and $i$ a natural number;
the unique factorization property of ideals in a Dedekind domain allows us to conclude that all $I$ and $J$ satisfy $N(IJ) = N(I) N(J)$.
% We also show that the norm of the principal ideal $\langle a \rangle$ in a number ring is equal to $|a|^n$, where $n$ is the degree of the ring.

We compute that the norm of \lstinline{sqrt_2 d} is the prime number $2$,
and conclude \lstinline{sqrt_2 d} itself is prime:
if it factors as $IJ$ then $N(IJ) = N(I) N(J) = 2$,
implying without loss of generality $N(I) = |R/I| = 1$ so $I = R$.

We conclude that, for $d\in \{-1,-2,-5,-6\}$, the class group of $\ZZ[\sqrt{d}]$ is generated by the class containing \lstinline{sqrt_2 d}, which we call \lstinline{class_group.sqrt_2 d}.
The fact that this ideal squares to two shows that this class has order $1$ or $2$. In particular, the class group consists of at most $2$ elements:
\begin{lstlisting}
theorem class_group_eq (d : ℤ)
  [fact (d % 4 = 2 ∨ d % 4 = 3)]
  [fact (squarefree d)]
  (I : class_group ℤ[√d]) :
  -6 ≤ d → d ≤ 0 → I ∈ {1, sqrt_2 d}
\end{lstlisting}

Similarly, but with more effort, we show that the class group of $\ZZ[\sqrt{-13}]$ is generated by these two classes. We take the set $M=\{1,2,3,4\}$ for the norm estimates ($\{1,2,3\}$ would also work) and appeal to the Kummer-Dedekind theorem for showing that $\langle 3 \rangle$ is a prime ideal in $\ZZ[\sqrt{-13}]$.

For applications to solving the Mordell equations for the $d$'s covered by this theorem this is actually enough information to complete the proof, as for those applications all that is required is that $3$ does not divide the class number.
Nevertheless, in the general case it will be necessary to compute the class number fully rather than just an upper bound of 2,
so we also formalize methods to exhibit distinct elements of the class group and give nontrivial lower bounds on the size.

\subsection{Lower bound}

In order to complete our computation of the class groups,
we need to establish whether the classes we found are distinct.
For this, we need to establish whether or not certain ideals are principal or not.
While we can easily certify principality by giving an explicit generator, proving non-principality may be more tricky.
For our examples however, we can quickly get the necessary proofs of non-principality in quite some generality, at least when $d < -2$.
Together with the already established principality of \lstinline{(sqrt_2 d)^2} we get:

\begin{lstlisting}
lemma order_of_sqrt2 (d : ℤ)
  [hd : fact (d % 4 = 2 ∨ d % 4 = 3)] [fact (squarefree d)] :
  d < -2 → order_of (class_group.sqrt_2 d) = 2
\end{lstlisting}

To prove \lstinline{sqrt_2 d} is not principal, we take the norm of a hypothetical generator $a + b\sqrt{d}$, giving an equation $a^2 - d b^2 = 2$ with $a, b \in \ZZ$.
A simple estimate shows that there cannot be any solutions for $d < -2$.
Lagrange's theorem then implies:
\begin{lstlisting}
theorem two_dvd_class_number (d : ℤ) (d' : ℚ)
  [fact (algebra_map ℤ ℚ d = d')] [fact (d % 4 = 2 ∨ d % 4 = 3)] [fact (squarefree d)] :
  d < -2 → 2 | class_number ℚ(√d')
\end{lstlisting}
This says that $2 \mid h(\QQ(\sqrt{d}))$ for any squarefree integer $d<-2$ with $d \equiv 2 \text{ or } 3 \pmod{4}$.

Originally the definition of class group in \mathlib depended on a choice of field of fractions,
meaning in the original version of our formalization we had to maintain a distinction between \lstinline{d : ℤ} and \lstinline{d' : ℚ}.
This made us repeatedly transfer hypotheses between \lstinline{d} and \lstinline{d'}.
Since every choice of field of fractions gives an isomorphic class group,
we redefined \lstinline{class_group} to instead use a specific construction \lstinline{fraction_ring} for the field of fractions,
and insert explicit isomorphisms when another choice is required.
This meant we only needed to transfer to \lstinline{d' : ℚ} at the last step of computing the class number.

\subsection{Explicit results}\label{sec:class_group_expl}

For $d\in\{-5,-6,-13\}$ the upper and lower bounds together imply that $h(\QQ(\sqrt{d}))=2$:
\begin{lstlisting}
lemma class_number_5 : class_number ℚ(√-5) = 2
lemma class_number_6 : class_number ℚ(√-6) = 2
lemma class_number_13 : class_number ℚ(√-13) = 2
\end{lstlisting}

In fact, we also get explicit (distinct) representatives for elements of the class group $\Cl(\QQ(\sqrt{d}))$ in these cases. For $d \in \{-1,-2\}$ it is not hard to show that \lstinline{sqrt_2 d} is principal, leading to $h(\QQ(\sqrt{d}))=1$:
\begin{lstlisting}
lemma class_number_1 : class_number ℚ(√-1) = 1
lemma class_number_2 : class_number ℚ(√-2) = 1
\end{lstlisting}

Alternatively, one could get these UFD results more directly by performing the norm estimates with $M= \{1\}$.
%(i.e. showing that $\QQ(\sqrt{-1})$ and $\QQ(\sqrt{-2})$ are norm Euclidean fields).

%The generalized division with remainder property from Theorem~\ref{thm main class group divisibility} is known to hold with $M=\{1,2,3\}$ for $\OK=\ZZ[\sqrt{-13}]$.
%This has not been formalized yet, but we believe that we will manage to do this in the very near future.
%We have already formalized that it implies the class number computation for $\QQ(\sqrt{-13})$:
%
%\begin{lstlisting}
%lemma class_number_eq_13 :
%  class_number ℚ(√-13) = 2
%\end{lstlisting}

\subsection{Other possible approaches}\label{sec:OtherApproaches}

It is known that, via the so called Minkowski bound, the class group of a number field $K$ of degree $n$ over $\QQ$ is generated by the classes of prime ideals $P$ with
\begin{equation}\label{eqn:Minkowski}
\Norm(P) \leq \sqrt{|\Delta_K|}\left(\dfrac{4}{\pi}\right)^{r_2}\dfrac{n!}{n^n}
\end{equation}
%\begin{proposition}
%Let $K$ be a number field of degree $n$, of discriminant $\Delta$ and let $r_1$ (resp. $r_2$) be the number of real (resp. complex) embeddings of $K$. Then every class in the class group of $K$ contains and integral ideal $I$ of norm $N(I) \leq M_K$, where
%\begin{equation*}
%M_K = \sqrt{|\Delta|}\left(\dfrac{4}{\pi}\right)^{r_2}\dfrac{n!}{n^n}.
%\end{equation*}
%\end{proposition}
where $r_2$ denotes the number of conjugate pairs of complex, non-real, embeddings of $K \hookrightarrow \CC$ and $\Delta_K$ denotes the discriminant of $K$ (see e.g.\ \cite[Section I.5]{Neukirch}). In our case, for $K=\QQ(\sqrt{d})$ with $1\not=d \in \ZZ$ squarefree, we have the identities
\begin{equation*}
\Delta_K = \left\lbrace
\begin{array}{ll}
d & \text{if } d \equiv 1 \pmod 4, \\
4d & \text{if } d \equiv 2,3 \pmod 4.
\end{array}
\right.
\end{equation*}

The main result underlying the proof of~\eqref{eqn:Minkowski} is Minkowski's convex body theorem, which has been formalized previously in both simple and dependent type theory (such as in Isabelle/HOL \cite{Minkowskis_Theorem-AFP} and Lean\footnote{\url{https://github.com/leanprover-community/mathlib/pull/2819}}).

While both the approach outlined in this section, and the Minkowski bound, are applicable to number fields of any degree, there is another method for computing class groups, or at least class numbers, in the setting of quadratic fields. In the case of $K=\QQ(\sqrt{d})$, for $d$ a negative squarefree integer, the Analytic Class Number Formula \cite[Section VII.5]{Neukirch} leads to a formula for the class number as an explicit finite sum
\begin{equation*}
h(K) = \dfrac{w}{2\Delta_K} \sum_{a=1}^{|\Delta_K|-1} \left(\dfrac{\Delta_K}{a}\right) a,
\end{equation*}
where $w$ is the number of units of $K$, i.e. $2$ unless $d=-1$ or $-3$ (in which case $w=4,6$ respectively), and $\left(\dfrac{\Delta_K}{a}\right)$ denotes the Jacobi symbol.
The main difficulty with formalizing this result for our purposes instead is that the analytic results needed are quite far removed from the more algebraic results used in the rest of this project.

\section{Explicit solution of Mordell equations}\label{sec:Mordell}

The focus of this paper is on the formalization of the solution of certain Mordell equations \eqref{eqn:Mordell} via computation of the class group and (ideal) descent. But we start with a brief history of the Mordell equation itself.

In 1657, Fermat claimed in a letter that the only solutions to $y^2 = x^3 - 2$ over the positive integers are $(x,y)= (3, 5)$.
In 1914, Mordell \cite{mordell14} studied the general equation $y^2 = x^3 + d$ from two perspectives, conditions on $d$ for non-existence of any integer solutions, and a method for finding solutions for certain $d$ based on ideal descent.
This equation soon became known as the \emph{Mordell equation}.
In 1918, he~\cite{mordell1920} then proved an ineffective finiteness theorem: for each nonzero $d \in \ZZ$, the Mordell equation has finitely many integer solutions.
See also Mordell's book~\cite[Chapter 26]{mordell1969diophantine}.
%
%In 1929, Siegel \cite{} proved in general that any smooth projective algebraic curve of positive genus defined over a number field $K$ only has finitely many points with coordinates in the ring of integers of $K$. This means that a Diophantine equation given by a polynomial over $K$ of degree larger than $2$ and satisfying certain technical assumptions only has finitely many solutions in the ring of integers of $K$.
%
In 1967, Baker \cite{baker68} proved the following effective result: for each $d \in \ZZ$, $d \neq 0$, if $x,y \in \ZZ$ satisfy $y^2=x^3+d$, then
    \begin{equation}\label{eqn:BakerBoundMordell}
        \max\{|x|,|y|\} \leq e^{10^{10}|d|^{10000}}.
    \end{equation}
%He used Thue equations and his then recently developed technique (which earned him a Fields medal) on bounds for linear forms in logarithms of algebraic numbers.
%
%Building on this, in 1973, Stark \cite{Stark1973} proved that for any fixed $\epsilon >0$, there exists an effectively computable constant $C_\epsilon$ such that any solution $(x,y) \in \ZZ^2$ to the Mordell equation satisfies
%\begin{equation*}
%    \max\{|x|,|y|\} \leq C_\epsilon^{|d|^{1+\epsilon}}.
%\end{equation*}
%
%In 1998, Gebel, Peth\"o and Zimmer \cite{} use their algorithm to compute integral points on elliptic curves to determine all solutions to \eqref{eqn:Mordell} for $0 < |d| < 10^4$ and for some $d$ with $10^4<|d|<10^5$.
Building on this, asymptotically sharper effective bounds were obtained. Most notably, Stark \cite{Stark1973} proved in 1973 that for any $\epsilon >0$, there exists an effectively computable constant $C_\epsilon$ such that now $\max\{|x|,|y|\}<C_\epsilon^{|d|^{1+\epsilon}}$. Using such bounds directly to solve a Mordell equation for any given $d$ is impractical, but techniques have been developed to resolve Mordell equations efficiently such as Bennett and Ghadermanzi \cite{bennett_mordells_2015} solved~\eqref{eqn:Mordell} for all (nonzero) $|d|<10^7$ in 2015.

%    Stub's to be worked out \emph{NC}:
%    \begin{itemize}
    %\item
    % blablbla \emph{Descent}\\
    %    The classical example where descent is used was done by Fermat, showing that
    %    \begin{equation}
    %        x^4+y^4=z^4
    %    \end{equation}
    %    has no nonzero integers solutions. Which has actually been formalized in Lean (footnote).

    %A key step, for $n \in \NN$ and $a,b,c \in D:=\ZZ$ is the descent step
    %\begin{equation}\label{eqn:UFD_descent}
    %    \gcd(a,b)=1 \text{ and } ab=c^n \implies a=u_1 c_1^n \text{ and } b=u_2 c_2^n
    %\end{equation}
    %for certain $c_1, c_2 \in D$ and units $u_1, u_2 \in D^*$.

    %\item explain very basics of quadratic fields/rings. (see chapter 2 of ``Dedekind domains''; define down to earth: number field, quadratic field, quadratic rings, ring of integers -- briefly!! no dedekind domains, refer to other paper -- and unit groups); mention needs not be UFD \ldots class group (vert quick def+main properties actually used).
%    \item (Move?) Mordell equations in general, finiteness (by Mordell?), effective finiteness (Siegell?), efficient finiteness (old-new...) including role of computer algebra, other interesting remarks (including quadratic recipricity arguments; Ireland-Rosen?).
%    Include
%\end{itemize}

Formalizing any of the above effective bounds, or the mentioned resolution for $|d|<10^7$, requires a very large amount of background theory to be developed first.
As a step in the direction of solving Diophantine equations beyond using elementary number theory, we formalize a flexible existence result using class group descent that appears in Mordell \cite[Section 8]{mordell14}.

\begin{theorem}\label{thm:Mordell_basic}
    Let $d<0$ be a squarefree integer with $d\equiv 2 \text{ or } 3 \pmod{4}$ and assume that $3\nmid h(\ZZ[\sqrt{d}])$. Then the Mordell equation~\eqref{eqn:Mordell} has solutions if and only if $d=-3m^2 \pm 1$ for some $m\in \NN$, in which case $y = \pm m (3d+m^2)$.
\end{theorem}

Let us now quickly sketch the proof of this result:
\begin{proof}
Suppose we have a solution $(x,y)$ to the Mordell equation~\eqref{eqn:Mordell}.
Then by moving $d$ to the other side and factoring this polynomial over the ring $D\coloneqq \ZZ[\sqrt{d}]$ we get
\[(y+\sqrt{d})(y-\sqrt{d})=x^3.\]
Hence for the ideals $I$, $J$, and $L$ generated by $y+\sqrt{d}$, $y-\sqrt{d}$, and $x$ respectively, we have $IJ=L^3$.
In the setting above, the ideals $I$ and $J$ are coprime as their GCD divides both $2y$ and $2\sqrt{d}$; however, it can be shown that $y$ and $d$ are coprime and that $x$ is odd, so that the GCD of the ideals must be one.
Since $D$ is the ring of integers of the number field $\QQ(\sqrt{d})$, we can invoke~\eqref{eqn:Ideal_descent}, which in particular implies that $I=L_1^3$ for an ideal $L_1$ in $D$.
Note that by construction $I$ is principal, so together with the assumption $3\nmid h(\ZZ[\sqrt{d}])$, we get from~\eqref{eqn:principality_from_power} that $L_1$ is principal.
This translates into
$y+\sqrt{d}=u \alpha^3$ for some $\alpha \in D$ and unit $u \in D^*$.
As all units in $D$ are cubes, we can and will assume without loss of generality that $u=1$. Writing $\alpha=a+b\sqrt{d}$ we arrive at
\[y+\sqrt{d}=(a+b\sqrt{d})^3=a(a^2+3b^2d)+b(3a^2+b^2d)\sqrt{d}.\]
Comparing coefficients of $\sqrt{d}$ on the left and right hand side yields
\[1=b(3a^2+b^2d).\]
As this constitutes a factorization of $1$ in the (rational) integers $\ZZ$, we get (substituting $b^2=1$) that
\[b=3a^2+d=\pm 1.\]%\qedhere\]
The corresponding $y$ value can now be computed and the result  follows quickly.
\end{proof}

\subsection{Formalizing solution of Mordell equations}

We have formalized the proof of Theorem~\ref{thm:Mordell_basic}, following the above sketch as the following statement in Lean:
\begin{lstlisting}
theorem Mordell_d (d : ℤ) (d' : ℚ) (hd : d ≤ -1)
  (hdmod : d % 4 = 2 ∨ d % 4 = 3)
  (hdd' : algebra_map ℤ ℚ d = d')
  (hdsq : squarefree d) [fact (¬ is_square d')]
  (hcl : (3 : ℕ).gcd (class_number ℚ(√d'))) = 1)
  (x y : ℤ) (h_eqn : y ^ 2 = x ^ 3 + d) :
  ∃ m : ℤ, ↑y = m * (d * 3 + m ^ 2) ∧
    (d = 1 - 3 * m ^ 2 ∨ d = - 1 - 3 * m ^ 2)
\end{lstlisting}

Note that we present an assumption to the typeclass system using \lstinline{fact} as in Section \ref{sec:QuadraticRings}.

The proof broadly follows the informal sketch just outlined.
First we factor the term $y^2 - d = (y + \sqrt{d})(y - \sqrt d)$, using arithmetic tactics \lstinline{ring} and \lstinline{calc_tac} to check the factorization.
Then we pass to an equality in \lstinline{class_group ℤ[√d] ℚ(√d')} and apply the ideal descent lemma (\ref{eqn:Ideal_descent}) to obtain that the ideal $\langle y + \sqrt{d} \rangle$ is a cube.
To check the side condition that the ideals $\langle y + \sqrt{d} \rangle,\langle y - \sqrt{d} \rangle$ are coprime requires many lines of code, and seems rather more complicated than informal presentations of these proofs. This is a situation where better tactics or abstractions could be of use.
The statement to be checked is the following:
\begin{lstlisting}
lemma gcd_lemma {x y d : ℤ} [fact (d % 4 = 2 ∨ d % 4 = 3)] [fact (squarefree d)]
  (h_eqn : y ^ 2 - d = x ^ 3) :
  gcd (span {⟨y, 1⟩}) (span {⟨y, -1⟩}) = 1
\end{lstlisting}
In pen and paper proofs authors generally take the contrapositive, and assume existence of a prime ideal $\mathfrak p$ dividing both $\langle y + \sqrt{d} \rangle,\langle y - \sqrt{d} \rangle$ and derive a contradiction with $\mathfrak p$ being prime.
However, for formalization we found it more convenient to simply compute the GCD of the two ideals by placing increasingly many restrictions on its norm.
%One reason for this convenience is the availability of tactics such as \lstinline{norm_num} for normalizing numerical expressions, this means we can calculate GCDs and divisibility of integers without any manual work.

One of the important ingredients of the \lstinline{gcd_lemma} is the fact that for the \lstinline{d} satisfying the assumptions of the theorem any solution to Mordell's equation necessarily has $x$ odd.
The strongest relevant version of this result is in fact:
\begin{lstlisting}
lemma x_odd_two_three_aux {x y d : ℤ}
  (hd : ↑d ∈ ({2, 3, 5, 6, 7} : finset (zmod 8)))
  (h_eqn : y ^ 2 - d = x ^ 3) : ¬ even x
\end{lstlisting}
This can be proved by generalizing to a statement about elements of \lstinline{zmod 8} and applying the built in decision procedure \lstinline{dec_trivial} to check all $8 \times 8 \times 8$ cases.
Note that this lemma also holds for $d \equiv 5 \pmod 8$, which is therefore the most similar case where the methods formalized here may apply when $d \not\equiv 2,3\pmod 4$.

From this we obtain an ideal whose cube is $\langle y + \sqrt{d} \rangle$, and which is principal by the assumptions on the class group.
Passing from an equality of principal ideals to their generators up to units is the application of very general lemmas in commutative algebra that hold in any domain.

The calculation of the units of $\ZZ[\sqrt d]$ for $d \le -1$ and $d\equiv 2,3\pmod 4$ follows quickly from the fact that the norm
$$\Norm(a+b \sqrt d) = a^2 - d b^2$$
can only equal 1 for small values of $a,b,d$ and can be handled using tactics for linear arithmetic (e.g. \lstinline{linarith}).
In all cases we can check that the unit group has order a power of two, and hence all units are cubes.

We conclude that
\lstinline{∃ z : ℤ[√d], z^3 = ⟨y, 1⟩}
which by casing on the components \lstinline{m, n} of \lstinline{z} and applying extensionality for \lstinline{quad_ring} results in the desired equalities of integers
\begin{lstlisting}
(m^2 + 3 * d * n^2) * m = y
(d * n^2 + 3 * m^2) * n = 1
\end{lstlisting}
which after factoring to find that $n = \pm 1$ give the witness for the existential in the theorem statement.

The converse to this theorem, showing that the given value of $y$ actually gives a solution when the condition that $d = -3m^2 \pm 1$ holds for some $m$, in fact holds in much more generality, and for far simpler reasons.
So we state and prove this as follows, simply checking that both cases give solutions to the equation over an arbitrary commutative ring:
\begin{lstlisting}
lemma Mordell_contra (d m : R)
  (hd : d = 1 - 3 * m^2 ∨ d = - 1 - 3 * m^2) :
  (m * (d * 3 + m^2))^2 = (m^2 - d)^3 + d :=
by rcases hd with rfl | rfl; ring
\end{lstlisting}

Now we come to applying \lstinline{Mordell_d} in specific cases.
Thanks to the calculation of the class number in Section \ref{sec:class_group_expl} we can apply the above result for $d \in \{-1,-2,-5,-6,-13\}$ unconditionally.
This gives the following theorems:
\begin{lstlisting}
theorem Mordell_minus1 (x y : ℤ)
  (h_eqn : y^2 = x^3 - 1) : y = 0
theorem Mordell_minus2 (x y : ℤ)
  (h_eqn : y^2 = x^3 - 2) : y = 5 ∨ y = -5
theorem Mordell_minus13 (x y : ℤ)
  (h_eqn : y^2 = x^3 - 13) : y = 70 ∨ y = -70
theorem Mordell_minus5 (x y : ℤ) : y^2 ≠ x^3 - 5
theorem Mordell_minus6 (x y : ℤ) : y^2 ≠ x^3 - 6

\end{lstlisting}

Note that in the first three cases the equation $d = - 3m^2 \pm 1$ has a solution, with $m=0$, $m = 1$, and $m=2$ respectively.
These give rise to the values of $y$ shown, which do give solutions of \ref{eqn:Mordell}.
In the last two cases no solution exists.

In both the non-existence results and the other cases of $d = - 3m^2 \pm 1$ that have no solution we need to show that certain quadratic equations in the integers have no solution.
This reduces to the checking whether specific rational numbers are squares.
While it would not be very difficult to write a decision procedure for this,
we opted to take a shortcut and simply check that the necessary quadratics have no solutions modulo a well chosen modulus $n$.
The advantage of this is that the ring \lstinline{zmod n} is already shown to have decidable equality and be finite in Lean's \mathlib, so that the built in decision procedure \lstinline{dec_trivial} will complete such goals with no extra work.
The downside of this approach is that the user must first find a modulus $n$ such that the equation considered has no solution by hand.

\subsection{Elementary and elliptic curve approaches}

As with the class group calculations, it seems fitting to discuss some other possible approaches to solving (certain) Mordell equations.
Most importantly, it will put the Mordell equations formally solved into some perspective.

%\subsection{Simple proofs of non-existence}
It is possible to give simpler proofs of the non-existence of integral solutions in some cases when none exist, by using less advanced number theoretic tools, such as quadratic reciprocity, unique factorization of integers, and congruences~\cite[Chapter XXIII]{dickson_history_2}.
%\footnote{see also \url{https://kconrad.math.uconn.edu/blurbs/gradnumthy/mordelleqn1.pdf}}.
These proofs simpler to formalize, as the technical prerequisites are lesser, but do not generalize to include more advanced cases where there are solutions.
% such as in the examples above (especially the $d=-13$ case).
    %In their 1990 textbook, Ireland and Rosen \cite{} state the following non-existence result use quadratic reciprocity to prove that, if \nc{find statement} then Equation \eqref{eqn:Mordell} has no integer solutions.

%\subsection{Elliptic curves}
Equation \eqref{eqn:Mordell} defines an \emph{elliptic curve} when viewed as an algebraic curve in the $(x,y)$ plane.
Mordell showed that the number of points with integer coordinates on an any elliptic curve (i.e. integer solutions to the defining equation) is finite \cite{mordell23}.
On the other hand, the rational points on an elliptic curve form an abelian group, called the Mordell-Weil group of the curve, which is finitely generated, often with generators of infinite order.
Determining the rank of an elliptic curve in general is an open problem, but currently it is possible to efficiently compute it in many cases.
%Computer algebra systems such as Magma, SageMath, and Pari/GP have good implementations of this calculation.
%The torsion part of the Mordell-Weil group can always be determined explicitly, via the theorem of Nagell and Lutz \cite{silverman_arithmetic_2013}.
%, we know that the torsion points of an elliptic curve have integer coordinates $(x,y)$ with $y^2$ dividing the \emph{discriminant} of the curve. In the case of a Mordell elliptic curve, the discriminant is given by $-2^4 \cdot 3^3 \cdot d^2$.
%In particular, if \eqref{eqn:Mordell} defines a curve of rank $0$, all its points are torsion points, thus the theorem of Nagell and Lutz allows one to quickly determine all (integer) solutions to it.
For Mordell curves it is known precisely when torsion can occur. %, and when $d$ is not a perfect power the Mordell curve is torsion free.
Therefore the most interesting cases of the Mordell equation from the general perspective are those where the rank of the curve is at least one, so that there are infinitely many rational solutions, but only finitely many integral solutions.

In the following table we list, for each integer $d$ with $-1 \geq d \geq -13$, the aforementioned quantities related to the Mordell elliptic curve $E_d : y^2 = x^3 + d$ corresponding to $d$, namely the rank $\rk(E_d)$ of the Mordell-Weil group of $E_d$ and the number of integer points $|E_d(\ZZ)|$ on it, as well as the class number $h(K_d)$ of the quadratic field $K_d=\QQ(\sqrt{d})$.
%All these are computed using Magma\ab{maybe we can just say any CAS can do this, or claim this data is "folklore"}.
% code for verification
% sage: for d in range(1,16):
% ....:     E = EllipticCurve([0,-d])
% ....:     print( QuadraticField(-d).class_number(),
% ....:      E.rank(), len(E.integral_points()))
% ....:
The bold elements in the table below denote the results that have been formalized.
%Notice that the case $d=-13$ is only formalized modulo the result mentioned in \S \ref{sec:class_group_expl}.
Moreover, the computation of the class number $h(K_d)$ for non-squarefree $d$ relies on the isomorphism with the quadratic field generated by the square root of the squarefree part of $d$.
%Explain the notation and say something about rank of elliptic curve and something about explicit torsion (NC)
\begin{center}
	\tabcolsep=0.15cm
	\begin{tabular}{c|c|c|c|c|c|c|c|c|c|c|c|c|c}
		$-d$          & 1                         & 2                         & 3 & 4 & 5                         & 6                         & 7 & 8 & 9 & 10 & 11 & 12 & 13                    \\ \hline
		$\rk(E_d)$     & 0                         & 1                         & 0 & 1 & 0                         & 0                         & 1 & 0 & 0 & 0  & 2  & 0  & 1              \\ \hline
		$|E_d(\ZZ)| $ & \textbf{1}                         & \textbf{2}                         & 0 & 4 & \textbf{0}                         & \textbf{0}                         & 4 & 1 & 0 & 0  & 4  & 0  & \textbf{2}\   \\ \hline
		$h(K_d)$      & \textbf{1}                         & \textbf{1}                         & 1 & \textbf{1} & \textbf{2}                         & \textbf{2}                         & 1 & \textbf{1} & \textbf{1} & 2  & 1  & 1  & \textbf{2}            \\
	\end{tabular}
\end{center}
For solving~\eqref{eqn:Mordell}, elementary non-existence proofs are obviously not available when $|E_d(\ZZ)|\not=0$, direct elliptic curve curve proofs are not available when $\rk(E_d)\not=0$, and in our approach the class group is unavoidable once $h(K_d)\not=1$.
%We note that the formally solved case 
Note that the case $d=-13$ has all these three characteristics.
%and therefore constitutes a quite nontrivial example of a formally solved Diophantine equation.

%\section{Sageify}
%As seen above, the solution of Mordell equations and computation of class groups involves many calculations in number fields and rings.
%For instance to verify whether two ideals are inverse in the ideal class group we must expand out several products of their generators, and check that one divides all the others typically.

%Reasons to select Sage:
%\begin{itemize}
%    \item It is open source and so available for free on many platforms
%    \item It unifies other mathematical software into a consistent interface
%    \item Interaction via a REPL is easy
%    \item Large and active community of developers means more functionality
%\end{itemize}

%Some downsides of the current setup
%\begin{itemize}
%    \item Slow to load
%    \item Brittle with respect to version changes
%\end{itemize}

\section{Computational tactics} \label{sec:computational-tactics}

In order to implement computations within \lstinline{quad_ring} we used a pair of tactics
that we called \lstinline{quad_ring.calc_tac} and \lstinline{times_table_tac}.
We initially used the tactic \lstinline{calc_tac} which is specific to \lstinline{quad_ring} and needs more manual guidance.
For more complicated goals, we implemented the more powerful and general tactic \lstinline{times_table_tac}.

\subsection{\lstinline{quad_ring.calc_tac}}

Taking advantage of our setup, the tactic \lstinline{calc_tac} can have an extremely short implementation in terms of existing \mathlib tactics:
\begin{lstlisting}
meta def calc_tac : tactic unit :=
`[rw quad_ring.ext_iff;
  repeat { split; simp; ring }]
\end{lstlisting}
First it uses the extensionality lemma \lstinline{quad_ring.ext_iff} to turn an equation \lstinline{x = y},
where \lstinline{x y : quad_ring R a b} into the conjunction \lstinline{x.b1 = y.b1 ∧ x.b2 = y.b2}.
Then the \lstinline{split} tactic replaces a single goal containing a conjunction with a pair of goals for the two sides of a conjunction.
The \lstinline{simp} tactic invokes the simplifier to rewrite the goal.
Finally, the \lstinline{ring} tactic (based on the Coq tactic of the same name~\cite{ring-tactic}) solves equalities of polynomial expressions by rewriting both sides of the equality in Horner normal form.
Thus, \lstinline{calc_tac} can solve any goal consisting of an equality in \lstinline{quad_ring}, such that the simplifier can turn the goal into componentwise equalities of polynomial expressions in commutative (semi)rings.

For example, to instantiate the \lstinline{comm_ring (quad_ring R a b)} structure, we have to provide a value for the field \lstinline{mul_comm (x y : quad_ring R a b) : x * y = y * x}.\\
We accomplish this by a single call to \lstinline{calc_tac}.
After applying extensionality and splitting the conjunction, the tactic produces two goals \lstinline{(x * y).b1 = (y * x).b1} and \lstinline{(x *} \lstinline{y).b2 = (y * x).b2}. % line break hack
We have previously registered \lstinline{@[simp]} lemmas specifying the components of a product:
\begin{lstlisting}
@[simp] lemma mul_b1 (z w : quad_ring F a b) :
  (z * w).b1 = z.b1 * w.b1 + z.b2 * w.b2 * b :=
rfl
@[simp] lemma mul_b2 (z w : quad_ring F a b) :
  (z * w).b2 = z.b2 * w.b1 + z.b1 * w.b2 +
    z.b2 * w.b2 * a := rfl
\end{lstlisting}
As the simplifier traverses the subexpressions of the goal, starting at the leaves, it uses the head symbol of the subexpression to look up all corresponding lemmas tagged \lstinline{@[simp]}, until matching succeeds.
It performs the rewrite and restarts its traversal in the new subexpression,
until the whole expression has been exhaustively rewritten.
In this case, the simplifier applies \lstinline{mul_b1} and \lstinline{mul_b2} twice each: once on the left hand side and once on the right hand side of the two goals.
The result is two goals \lstinline{x.b1 * y.b1 + x.b2 *} % evil line break hack
\lstinline{y.b2 * b = y.b1 * x.b1 + y.b2 * x.b2 * b} and \lstinline{x.b2 *} \lstinline{y.b1 +} % evil line break hack
\lstinline{x.b1 * y.b2 + x.b2 * y.b2 * a}.
Finally, the \lstinline{ring} tactic is called on both goals,
treating each side as multivariate polynomials in the variables \lstinline{x.b1 y.b1 x.b2 y.b2 a b}
and using commutativity of the ring \lstinline{R} to rewrite both sides into equal polynomials,
concluding the proof.

It is clear the applicability of \lstinline{calc_tac} rests fundamentally on the ability of the simplifier to reduce the components of an element of \lstinline{quad_ring} to a polynomial expression.
In other words, to be able to apply \lstinline{calc_tac}, the user has to curate a set of \lstinline{@[simp]} lemmas that are powerful enough to express all computations beyond multivariate polynomials.
Thus we accompanied every definition in \lstinline{quad_ring} with a pair of lemmas specifying its \lstinline{b1} and \lstinline{b2} components;
this can often be automated by applying the \lstinline{@[simps]} attribute to the definition.
On the other hand, the simplifier has to try all lemmas until it finds one that applies since it does no indexing of the set beyond the head symbol, so it slows down noticeably as the set of \lstinline{@[simps]} lemmas grows.
This concern is a relevant factor but not overriding, especially since Lean 4 promises a more efficient simplifier using discrimination trees~\cite{lean-4}.

Another disadvantage of the structure of \lstinline{calc_tac} is the potential to accumulate very large expressions during the simplification stage.
For example, the \lstinline{ring} tactic takes tens of milliseconds on the goal \lstinline{(x - y) * (x^2 + x * y + y^2) = x^3 - y^3},
while \lstinline{calc_tac} requires multiple seconds to prove the same equality over \lstinline{quad_ring},
since it first has to unfold every multiplication into a sum of multiplications, giving an intermediate result with 72 occurrences of variables.
Due to these limitations, we could apply \lstinline{calc_tac} only for simple equations such as the ring axioms for \lstinline{quad_}\- % ArXiv hyphenation hint
\lstinline{ring}.
In order to deal with more complicated equations in a more principled way, we developed a prototype of a more powerful tactic called \lstinline{times_table_tac}.

\subsection{Times tables}

The goal of \lstinline{times_table_tac} is to normalize expressions in a finite extension of a commutative ring, including but not limited to \lstinline{quad_ring}.
More precisely, let $S/R$ be an extension of commutative rings, such that $S$ is free and finitely generated as an $R$-module,
then \lstinline{times_table_tac} normalizes expressions in the ring $S[X_1, \dots, X_n]$ of multivariate polynomials given a normalization procedure for expressions in $R$.

Finite ring extensions suitable for \lstinline{times_table_tac} occur commonly in mathematics,
a particularly relevant example for our paper being $S = R[\alpha]$ where $\alpha$ is a root of a monic polynomial,
since \lstinline{quad_ring} is the special case of $R[\alpha]$ where $\alpha$ has degree two.
Further common examples of ring extensions that satisfy the preconditions required by \lstinline{times_table_tac} include
the complex numbers over the real numbers,
all matrix rings $M_{n}(R)$ over a given ring $R$,
and all finite fields $GF(p^k)$ over $\ZZ/p\ZZ$.

We can define a normal form for $S$ by writing every $x \in S$ as an $R$-linear combination of a family of basis vectors $b$;
the coefficients of this linear combination are well-defined and unique if $S$ is a free $R$-module.
Addition and scalar multiplication of these linear combinations can be done componentwise,
but for multiplication of two elements of $S$ we need additional information.
Because a ring extension $S$ of $R$ is also an $R$-algebra, multiplication in $S$ is an $R$-bilinear map,
and if we are given a finite basis $b$ of $S$ as an $R$-module (which exists because we assume $S$ is free and finite),
then we can specify the multiplication operation by giving the product $T_{i, j} \in S$ of each pair of basis elements $b_i$, $b_j$.
Therefore, to multiply two elements $x y \in S$, we can write each as a linear combination of basis elements $x = \sum_i c_i b_i$ and $y = \sum_i d_i b_i$,
so by distributivity we find $xy = \sum_{i, j} T_{i, j} c_i d_j$.
Writing $T_{i, j} = \sum_k t_{i, j, k} b_k$, the normal form of $xy$ is given by $\sum_{i, j, k} c_i d_j t_{i, j, k} b_k$.
The matrix $T$ of products of basis elements is the source of the name \lstinline{times_table_tac}.

For example, if we set $S = \ZZ[\sqrt d]$, we can choose $\{1, \sqrt d\}$ as basis vectors,
with times table given by
\begin{align*}
	T = \begin{pmatrix}
		1 & \sqrt d \\
		\sqrt d & d
	\end{pmatrix}.
\end{align*}

In order to also support the use of variables in the expression,
we can construct a normal form for $S[X_1, \dots, X_n]$ if $S$ is an $R$-module of degree $d$,
by working in the polynomial ring $R[X_{1, 0}, \dots, X_{1, d}, X_{2, 0}, \dots, X_{n, d}]$
and setting $X_1 = \sum_i X_{1, i} b_i$.
For the ring of multivariate polynomials, a suitable normal form is the Horner normal form.

The implementation of \lstinline{times_table_tac} works bottom-up to construct a normal form for each subexpression,
evaluating sums, scalar multiples, and products of normal forms into a new normal form as it encounters them.
Normalizing during evaluation ensures terms are kept relatively small, avoiding the issues of \lstinline{calc_tac} on more complicated expressions.
The normal form is recorded as a vector of Horner normal forms, reusing the implementation of the \lstinline{ring} tactic to manipulate Horner expressions \cite{ring-tactic}.
In addition to computing a normal form, the tactic also assembles a proof that the input expression is equal to its normal form.
Constructing a proof term tends to be faster in Lean 3 than proof by reflection, since the interpreter for metaprograms is usually more efficient than kernel reduction.
The data of a basis and the associated times table is bundled in a structure \lstinline{times_table}:
\begin{lstlisting}
structure times_table (ι R S : Type*) [fintype ι]
  [semiring R] [add_comm_monoid S] [module R S]
  [has_mul S] :=
(basis : basis ι R S)
(table : ι → ι → ι → R)
(unfold_mul : ∀ x y k, repr basis (x * y) k =
  ∑ i j : ι, repr basis x i * repr basis y j * table i j k)
\end{lstlisting}
The fields \lstinline{basis} and \lstinline{table} correspond to the variables $b$ and $t$ above, respectively.
Here, \lstinline{repr basis} is a linear equivalence sending a vector \lstinline{x : S} to the family of coefficients \lstinline{c}
such that \lstinline{∑ i, c i • basis i = x}.

We provide a user interface for \lstinline{times_table_tac} in the form of an extension to the \lstinline{norm_num} tactic.
This tactic is intended to evaluate numeric expressions occurring in a goal or hypothesis, by rewriting them to an explicit numeral.
Spe\-ci\-fic\-al\-ly, the \lstinline{times_table_tac} extension evaluates expressions of the form \lstinline{repr (times_table.basis T) x}.
Thus, the times table is passed to the tactic by the expression we want to normalize.

Note that the definition of \lstinline{times_table} does not assume \lstinline{S} has a ring structure.
This allows us to use the tactic to prove the ring axioms
given an explicit definition of multiplication in terms of a times table,
in the same way that we used \lstinline{calc_tac} to construct the \lstinline{comm_ring} instance on \lstinline{quad_ring}.

Since \lstinline{quad_ring.calc_tac} proved sufficient for our current purposes,
we did not apply \lstinline{times_table_tac} in practice and we left it as a proof of concept.
From our test cases, we determined that \lstinline{calc_tac} takes noticeably more time than \lstinline{times_table_tac} as expression sizes grow:
on goals with few operations, such as commutativity or associativity of multiplication in $\QQ(\sqrt{d})$,
\lstinline{calc_tac} is approximately two times faster than \lstinline{times_table_tac},
while the ratio decreases to $\frac{3}{2}$ on larger goals such as $x^4 - y^4 = (x - y) (x^3 + x^2 y + x y^2 + y^3)$.
Still, this means that both tactics suffer from noticeable slowdowns as expressions grow: they need more than a minute to solve this relatively easy goal.
Unfortunately the normalization steps that reduce summations such as $\sum_{i, j, k} c_i d_j t_{i, j, k} b_k$ take up enough time that \lstinline{calc_tac} is faster on the small expressions we encountered in our development.
We believe that a good caching strategy can remedy these issues when implementing the full incarnation of \lstinline{times_table_tac},
since the same expressions $c_i$, $d_j$ and $t_{i, j, k}$ will tend to reoccur when unfolding the definition.

\section{Links to computer algebra systems}\label{sec:Sageify}

In much of the above work there is a lot of time spent on calculations that are routine, and can be done algorithmically.
In particular, many of these computations once performed can be certified more quickly than they can be carried out.
For instance, checking membership of elements in ideals, or more generally checking containments or equality between two ideals given by explicit lists of generators is a possibly nontrivial task, depending on the base ring, but can always be certified along as efficient arithmetic can be done in the ring.
This presents a choice to the formalizer, either to implement computationally efficient versions of such algorithms in the proof assistant they are to be used, or in its meta language, or to do these calculations externally in an existing computer algebra system and import and verify the results somehow.

Implementing these computations directly can be prohibitively difficult and time consuming.
Depending on the calculation needed, it may take many months to implement state of the art algorithms inside a proof assistant.
Importing computations from an external tool can be done somewhat manually, but can add significant overhead and reduce readability if the terms involved are large.

In order to automate the process of calling an external system mid-proof, we have developed a tactic that uses the SageMath computer algebra system~\cite{sagemath} as a backend to determine drop in replacement of data that is of a more useful form, either somehow simplified or broken up into pieces that make some property of the term more clear.
The main tactic \lstinline{replace_certified_sage_equality} will be described below.
It is intended to be a flexible way to allow users to add their own tactics using SageMath for unverified computations and then to verify them in only a few lines.
This system is similar in goals to \cite{lewis_extensible_2017, lewis_bi-directional_2022}, but we aim to make our system more widely usable in some respects.

The system is intended to be flexible so that it can be used to implement many tactics that fit this profile.
For instance, we can implement tactics to factor integers and polynomials, decompose into a divisible part and a remainder, or replacing a matrix by a change of basis matrix multiplied by a matrix in some normal form.
Details on how the access to SageMath has been implemented can be found in~\cref{app:access}.
We intend to use this paradigm in parts of our work that require explicit calculations, but have not yet done so.

\subsubsection*{Design}
\lstinline{replace_certified_sage_equality} has signature:

\begin{lstlisting}
meta def replace_certified_sage_equality
  (matcher : expr → tactic bool)
  (reify_type : Type*)
  (reify : expr → tactic reify_type)
  (sage_input : reify_type → tactic string)
  (output_type : Type*)
  [has_from_sage output_type]
  (convert_output : output_type → tactic expr)
  (validator : expr → expr → tactic (expr))
  (name : string)
  (wi : parse (tk "with" *> pexpr)?) :
  tactic unit
\end{lstlisting}
This \lstinline{replace_certified_sage_equality} tactic iterates over the goal, and finds subexpressions that \lstinline{matcher} returns true on.
The first one found is then reified into explicit data of type \lstinline{reify_type} using \lstinline{reify}; this means that even terms of types that are not computable, can be reified provided the \lstinline{reify} function can recognise the term.
The \lstinline{sage_input} function is then used to create a string, the command that is passed to SageMath.
The output of SageMath (via one of the backends) is read as a string which is parsed into \lstinline{output_type} (this parsing is handled via typeclasses to make it easy to parse types such as nested tuples and lists).
\lstinline{convert_}\-% ArXiv linebreak hack
\lstinline{output} is then responsible for constructing an expression representing a term of the same type as the originally matched input, the equality of these is then proved by \lstinline{validator} that returns a proof of equality. This is then rewritten to replace the originally matched expression with the validated version returned by SageMath.
The final two arguments are used to parse an optional \lstinline{with "a string"} after the tactic, which is used in the self replaced version.

Using this boilerplate tactic, a tactic to factor natural numbers using SageMath can be implemented in only about 20 lines (see \cref{app:factor_nats}), without the implementer needing to know the details of how SageMath is invoked.
For this example we reify expressions representing naturals to naturals themselves and ask SageMath for a factorization as a list of pairs of prime factors and exponents. We then construct the corresponding factored term as a product of powers.
The \mathlib tactic \lstinline{norm_num} is used to prove the equality of the original expression and the resulting one.

\section{Discussion}

\subsection{Future work}\label{sec:Future}

Apart from the approached mentioned in Section~\ref{sec:OtherApproaches}, especially using Minkowski's bound, there are several natural ways in which the present work can be expanded on.

Modifications could be made to consider other Mordell equations, including the following.
%Finishing the case when $d=-13$ would provide a very interesting example.
%
%When $d = -13$, by generalizing the class number bounds above to show that the class group is generated by primes above 2 and 3, and then applying the Kummer-Dedekind theorem to show that 3 is inert, once the class group is shown to be of order 2 the results formalized here can be directly applied to complete the proof.
%
When $d = -7$, this would involve considering discriminants congruent to 1 mod 4, and therefore working over \lstinline{quad_ring} % evil line break hack
\lstinline{ℤ 1 ((d - 1) / 4)}. This should not be significantly more complicated than the proofs above. % TODO check
When $d=-23$, this would involve dealing with $h(K)=3$. For this we would need to use explicit ideal representatives for the elements of the class group (which we would compute in principle).
%
%When $d = 2$, although this appears quite different to the cases considered so far as $d$ is positive, there exists a change of variables technique \cite[p. 399]{uspensky_elementary_1939}, which instead works over $\ZZ[\sqrt{-6}]$ and should be formalizable without too much additional work.%\footnote{\url{https://kconrad.math.uconn.edu/blurbs/gradnumthy/mordelleqn2.pdf}}.\ab{cite some number theory book from 1939 instead}
%
When $d > 0$, this would involve more work as the unit group of the ring of integers of $\QQ(\sqrt{d})$ (assuming $d$ is not a square) is then infinite. % by Dirichlet's unit theorem.
There will still be finitely many possibilities for the units modulo cubes, but this will require some additional casework when passing from principal ideals to elements, in addition to the problem of calculating generators for the unit group.

We could also consider more general forms of equations.
The above discussed techniques can still be of use to solve equations more general than the Mordell equation.
For instance, the exponents $2,3$ in $y^2 = x^3 +d$ could be changed, or even more generally, superelliptic equations of the form $y^d = f(x)$ for an arbitrary polynomial $f \in \QQ[x]$ may be considered.
This involves working in the algebra $\QQ[x]/\langle f\rangle$ and would require a significant advance in verified algorithms for computing class and unit groups at least.
%Nevertheless, this is one situation where the casework to be considered to successfully solve such an equation over the integers is difficult to carry out by hand.
    %In this situation the \lstinline{sagify} tactic, or some other connection to a computer algebra system to do as much work as possible seems preferable, rather than reimplementing entire libraries of number theoretic algorithms from scratch.

On the side of tactics it is clear that many goals arising when working with explicit number fields and ideals will be of fairly specific forms, checking some equality or calculation in a number field, checking equality of two ideals, or finding normalized sets of generators for them, checking equivalence of ideals in the class group, or showing non-principality of a given ideal.
Specialized tactics for solving goals like these should be within reach, and will massively speed up progress when formalizing arguments in this field.

One application of solving Mordell equations is being able to determine all elliptic curves over $\QQ$ with prescribed bad reduction.
This could also be formalized in future work.

\subsection{Related work}\label{sec:Related}

Previous work on Diophantine equations includes mostly elementary resolutions, such as work of Fermat, Euler, Gauss, and Legendre.
There are several formalizations of this type of material, which involves explicit change of variables, checking for solutions modulo primes, quadratic reciprocity, and descents within $\ZZ$.
For example, the exponent 4 case of Fermat's Last Theorem has been formalised in several proof assistants, including Isabelle/HOL (see~\cite{Oosterhuis}, also for the exponent 3 case), Coq (see~\cite{Diophante-Fermat}), and Lean\footnote{\href{https://github.com/leanprover-community/mathlib/blob/4e1eeebe/src/number\_theory/fermat4.lean}{https://github.com/leanprover-community/mathlib/blob/4e1eeebe/src/num}\\
\href{https://github.com/leanprover-community/mathlib/blob/4e1eeebe/src/number\_theory/fermat4.lean}{ber\_theory/fermat4.lean}}.
%\sd{If more space is needed: move (long) mathlib link to bibliography}
%\ab{maybe cite or link to \url{https://fse.studenttheses.ub.rug.nl/8392/1/Roelof_Oosterhuis_doctoraal.pdf}, to back up the claim this sort of thing has been done for a long time, but nobody went further, this thesis quite explicitly says one needs class groups to go further and theory is needed}

There have been several formalizations concerning Diophantine equations and problems from the perspective of mathematical logic (for example see \cite{larcheywendling_et_al} or \cite{pak_et_al} and the references therein).
This area has some overlap with the basic objects of study and problems considered here, but in terms of methodology and goals remains quite distinct.
%In the present work we consider equations with a simple form, two variables and small coefficients and formalize number theoretic tools to find the set of solutions very explicitly, when it is not clear from the outset whether solutions exist or not.
%In the study of Diophantine equations from the perspective of logic, it is instead more interesting to construct Diophantine equations whose solutions describe some set of tuples of numbers with a specified property.

The solution of the Pell equation ($x^2 -dy^2 =1$) is one of the `top 100' mathematical theorems listed as formalization targets tracked by Freek Wiedijk, and some aspects of this equation have been formalized in HOL Light, Isabelle/HOL, Metamath, Lean, and Mizar.
This is related to computing the group of units in real quadratic fields, which plays a role when solving a Mordell equation for positive $d$.
The formalization in Lean's \mathlib however is specialized to the case of $d = a^ 2 - 1$ for some $a \in \ZZ$.
This makes it only useful for a very small fraction of the real quadratic fields that arise when solving Mordell equations, and so without generalization it is not as useful for Diophantine applications as a formalization as it could be.
%Abel-Ruffini \cite{bernard_unsolvability_nodate}

There are initiatives towards formaliziations connected to Fermat's Last Theorem, most notably the ongoing \lq FLT regular\rq\ project\footnote{\url{https://github.com/leanprover-community/flt-regular}}, %\sd{If more space is needed: move FLT regular link to bibliography}
aiming to formalize Kummer's 19th century work on Fermat's Last Theorem for exponents which are so-called regular primes.

De Frutos-Fernández has formalized the theory of adèles and idèles of number fields \cite{de_frutos-fernandez_formalizing_2022}, which provides another more theoretical way of thinking about the ideal class group, and begins the development of class field theory.

\subsection{Conclusion}\label{sec:Conclusion}

We have successfully formalized the solutions to several Mordell equations,
and more importantly built theory and tactics that can be used for solving other instances of the Mordell equation or Diophantine equations in general.
We found that a large part of our efforts went into working out every detail of computations such as equality of ideals that would usually be left to a computer algebra system, we prototyped the design of some tactics to make this easier in future.
On the other hand, more theoretical arguments such as primality of the ideal \lstinline{sqrt_2 d} from primality of its norm formalized smoothly.

In total, our development took up about 6000 lines of code, of which about 1000 lines concern preliminaries, and the rest are approximately equally distributed over Sections \ref{sec:QuadraticRings}---\ref{sec:computational-tactics} of this paper. %\tb{Checked 2022-09-20, 20:00 UTC using cloc}
The De Bruijn factor for a project such as this is very hard to measure reliably as a whole.
This is as the results formalized here have many prerequisites both on paper and in a formal system, there is no single book or article that starts from close to no mathematical background and presents the results formalized here.
As such, any measure of the De Bruijn factor depends heavily on what the authors choose to consider as background material or as implicit in informal works.

\begin{acks}
Anne Baanen was funded by NWO Vidi grant No. 016.Vidi.\linebreak[0]189.037, Lean Forward.
Alex J. Best, Nirvana Coppola, and Sander R. Dahmen were funded by NWO Vidi grant No. 639.032.613, New Diophantine Directions.

We would like to thank Jasmin Blanchette for his comments on an earlier version of the paper.
We thank the anonymous referees for their helpful comments.
Finally, we would like to thank the \mathlib community for their support throughout the development process.
\end{acks}

\appendix

\section{Appendix: Access to systems}\label{app:access}

We believe that using a free and open source computer algebra system as a backend helps ensure that the tactics will be potentially useable my many users.
In order to increase the likelihood that such a tactic is usable on a particular system without additional setup we have implemented several backends for calling SageMath on the user's machine directly, or installed via conda\footnote{\url{https://docs.conda.io}}.
In case the user does not have SageMath installed on their machine we have added three backends to access a publicly accessible SageMath instance over the internet, via \url{https://sagecell.sagemath.org}, by using one of curl, ncat or openssl.
The latter approach was previously used by the \mathlib tactic \lstinline{polyrith}, written by Dhruv Bhatia.
These are the main reasons we choose to use SageMath for these tactics, though it should be simple  to use other systems in its place without large changes as the interface is via plaintext.

In addition to ensuring that a variety of ways of accessing SageMath can be used, we also designed this tactic in such a way that it is self-replacing.
As SageMath returns string outputs, once the tactic is run once, future uses of it can be fed the appropriate SageMath output directly.

For example, using a tactic to factor natural numbers implemented via this schema, the user is recommended to replace the call to the tactic with one containing the explicit output of SageMath:
\begin{lstlisting}
example : ¬ nat.prime 1111 := begin
  factor_nats, -- Try this: factor_nats with "[(11, 1), (101, 1)]"
  simp [nat.prime_mul_iff], -- a product can only be prime if all factors are one
end
\end{lstlisting}

\section{Appendix: Example of a CAS certification tactic for factoring}
\label{app:factor_nats}
\begin{lstlisting}
meta def tactic.interactive.factor_nats := replace_certified_sage_equality
  (λ ex, do
    t ← infer_type ex,
    return $ t = `(nat))
  ℕ
  (λ e, e.to_nat)
  (λ n, return sformat!"print(list(ZZ({n}).factor()))")
  (list (ℕ × ℕ))
  (λ l, do
    ini ← l.mfoldl (λ ol ⟨p, n⟩, do
      P ← expr.of_nat `(ℕ) p,
      N ← expr.of_nat `(ℕ) n,
      return `((%%ol : ℕ) * (%%P : ℕ) ^ (%%N : ℕ))) `(1 : ℕ),
    sl ← simp_lemmas.mk_default,
    prod.fst <$> simplify sl [] ini)
  (λ o n, do
    (e₁', p₁) ← or_refl_conv norm_num.derive o,
    (e₂', p₂) ← or_refl_conv norm_num.derive n,
    is_def_eq e₁' e₂',
    mk_eq_symm p₂ >>= mk_eq_trans p₁)
  "factor_nats"
\end{lstlisting}

\bibliographystyle{ACM-Reference-Format}
%% TODO deduplicate the bib
\bibliography{MordellLean.bib}

\end{document}